\documentclass[11pt]{article}
\pdfoutput=1
\usepackage[centertags]{amsmath}
\usepackage[square, comma, sort&compress,numbers]{natbib}
\usepackage{array,multirow}
\numberwithin{equation}{section}
\usepackage{amssymb}
\usepackage{graphicx}
\usepackage{color}

\usepackage{epsfig}

\def\Or[#1]{{\text{O}}\left({#1}\right)}
\def\dotl[#1,#2]{\left\langle #1, #2 \right\rangle}
\def\dotlb[#1,#2]{[ #1, #2 ]}
\def\dotp[#1,#2]{(#1) \cdot (#2)}
\def\aff[#1,#2]{\hat{#1}(#2)}
\def\n4sym{{\cal N}=4 SYM}
\def\>{\rangle}
\def\<{\langle}
\def\weight[#1,#2,#3]{\{(#1),#2,#3\}}
\def\ads[#1]{$\text{AdS}_{#1}$}

\newcommand{\ba}{\begin{eqnarray}}
\newcommand{\ea}{\end{eqnarray}}
\newcommand{\be}{\begin{eqnarray}}
\newcommand{\ee}{\end{eqnarray}}
\newcommand{\bq}{\begin{equation}}
\newcommand{\eq}{\end{equation}}
\newcommand{\benn}{\begin{equation*}}
\newcommand{\eenn}{\end{equation*}}
\newcommand{\bi}{\begin{itemize}}  
\newcommand{\ei}{\end{itemize}}

\newcommand{\CO}{{\cal O}}

\newcommand{\nn}{\nonumber}

\newcommand\oo\infty
\newcommand\s\sigma
\newcommand\de\delta
\newcommand\De\Delta

\newcommand\f\phi
\newcommand\g\gamma
\newcommand\x\times

\newcommand{\ra}{\rightarrow}
\newcommand{\lra}{\leftrightarrow}
\newcommand{\fr}{\frac}

\newcommand{\Ocal}{{\cal O}}

\newcommand{\Ncal}{{\cal N}}

\usepackage{jheppub}

\makeatletter
\def\@fpheader{\vspace{-.1cm}}
\makeatother

\title{Hawking from Catalan}

\author[a,b,c]{A.\ Liam Fitzpatrick,}
\author[d]{Jared Kaplan,}
\author[c]{Matthew T.\ Walters,}
\author[d]{and Junpu Wang}
\affiliation[a]{Stanford Institute for Theoretical Physics, Stanford University, \\
Via Pueblo, Stanford, CA 94305, U.S.A.}
\affiliation[b]{SLAC National Accelerator Laboratory, \\
Sand Hill Road, Menlo Park, CA 94025, U.S.A.}
\affiliation[c]{Department of Physics, Boston University, \\
Commonwealth Avenue, Boston, MA 02215, U.S.A.}
\affiliation[d]{Department of Physics and Astronomy, Johns Hopkins University, \\
Charles Street, Baltimore, MD 21218, U.S.A.}
\emailAdd{fitzpatr@stanford.edu}
\emailAdd{jaredk@pha.jhu.edu}
\emailAdd{mtwalter@bu.edu}
\emailAdd{jwang217@jhu.edu}
\abstract{
The Virasoro algebra determines all `graviton' matrix elements in AdS$_3$/CFT$_2$. 
We study the explicit exchange of any number of Virasoro gravitons between heavy and light CFT$_2$ operators at large central charge.  These graviton exchanges can be written in terms of new on-shell tree diagrams, organized in a perturbative expansion in $h_H / c$, the heavy operator dimension divided by the central charge.  
The Virasoro vacuum conformal block, which is the sum of all the tree diagrams, obeys a differential recursion relation generalizing that of the Catalan numbers.  We use this recursion relation to sum the on-shell diagrams to all orders, computing the Virasoro vacuum block.  Extrapolating to large $h_H / c$ determines the Hawking temperature of a BTZ black hole in dual AdS$_3$ theories.
}
\arxivnumber{}

\begin{document}

\maketitle
\flushbottom

\section{Introduction}

Some features of AdS/CFT are universal, and can be derived using only quantum mechanics and conformal invariance \cite{FerraraOriginalBootstrap1,PolyakovOriginalBootstrap2, Cardy:1986ie, Rattazzi:2008pe, JP}.  Examples include locality at distances much larger than the AdS scale and long-range forces \cite{Fitzpatrick:2012yx, KomargodskiZhiboedov, Fitzpatrick:2014vua, Alday:2013cwa,  Alday:2014tsa, Alday:2015eya, Jackson:2014nla, Kaviraj:2015cxa, Kaviraj:2015xsa, Fitzpatrick:2015qma, Hellerman:2015nra}, asymptotic limits of partition functions \cite{Hellerman:2009bu, Hartman:2014oaa}, causality constraints \cite{Hartman:2015lfa}, and connections between entanglement entropy and geometry \cite{RT1, RT2, HartmanLargeC, Faulkner:2013yia, Faulkner:2014jva, Swingle:2014uza, Caputa:2015waa}.  Recently, it has been argued that at large central charge, features of AdS$_3$/CFT$_2$ thermodynamics become simple and universal \cite{Hartman:2014oaa, Fitzpatrick:2014vua, Roberts:2014ifa, Jackson:2014nla, Fitzpatrick:2015zha}.  

One expects that with an economical choice of dynamical assumptions on conformal theories, one can go further and single out Einstein gravity in AdS, thus implying a host of non-trivial predictions for the boundary theory.  Effective field theory arguments suggest that a large $N$ CFT where the only operators with dimensions below some large `gap'  are products of stress tensors should be well-described by the Einstein-Hilbert action in AdS, so that the full action can be fixed by just the graviton three-point function \cite{JP, ElShowk:2011ag, AdSfromCFT, Camanho:2014apa, Bellazzini:2015cra}.  On the CFT side, this suggests that such a theory  should have correlators that are fixed by the stress tensor three-point function.  This would be reminiscent of the construction of gravitational scattering amplitudes purely from the on-shell three-point function \cite{Benincasa:2007qj, ArkaniHamed:2008yf}.  More generally, some features of thermodynamics, particularly at high temperatures, may hold universally even in the absence of a gap in dimensions of operators. 

In $d>2$ dimensions, demonstrating such predictions explicitly is challenging, since general correlators of stress tensors are not determined from their three-point function.  However, as is well-known, CFTs in $d=2$ dimensions have an infinite Virasoro spacetime symmetry algebra that {\it does} fix general correlators of stress tensors in terms of their three-point function, or equivalently in terms of the central charge $c$ of the theory. In this note we demonstrate that simple `on-shell diagrammatic rules' emerge from the Virasoro algebra in a `heavy-light' limit, and computing correlators in this limit amounts to summing over diagrams. Each diagram takes the form of a `binary tree diagram' like Figure \ref{fig:VirasoroDiagramExample}.  The diagrammatic rules are simple enough that the problem of summing all diagrams can be related to the problem of counting the number of tree diagrams with $k$ vertices, i.e.\ the Catalan numbers $C_k$:
\be
C_k = \left\{ \textrm{number of distinct trees with $k$ vertices} \right\} .
\ee
The Hawking temperature of dual BTZ black holes arises as the analytic continuation of the generating function of the $C_k$.

The holomorphic Virasoro generators $L_n$ originate as the modes in an expansion of the holomorphic stress tensor\footnote{It is conventional \cite{Ginsparg} to define $T(z) \equiv -2 \pi T_{zz}(z)$.  It does not depend on the anti-holomorphic coordinate $\bar z$ because conservation requires $\partial_{\bar z} T_{zz} = 0$.  CFT$_2$ have two commuting copies of Virasoro, corresponding to the holomorphic and anti-holomorphic stress tensor; we will focus on the holomorphic sector, with identical results following for the independent anti-holomorphic sector.  See \cite{Ginsparg, Fitzpatrick:2014vua} for a more pedagogical discussion.}
\be
T(z) = \sum_n z^{-2-n} L_n.
\ee
The global conformal generators $L_{\pm 1}$ and $L_0$ are symmetries of the CFT vacuum, but the $L_{-m}$ with $m \geq 2$ are `dynamically broken' and do not annihilate the vacuum.  Instead, these generators act on the vacuum to create non-trivial states whose matrix elements are  determined by Virasoro symmetry.  

\begin{figure}[t!]
\begin{center}
\includegraphics[width=0.45\textwidth]{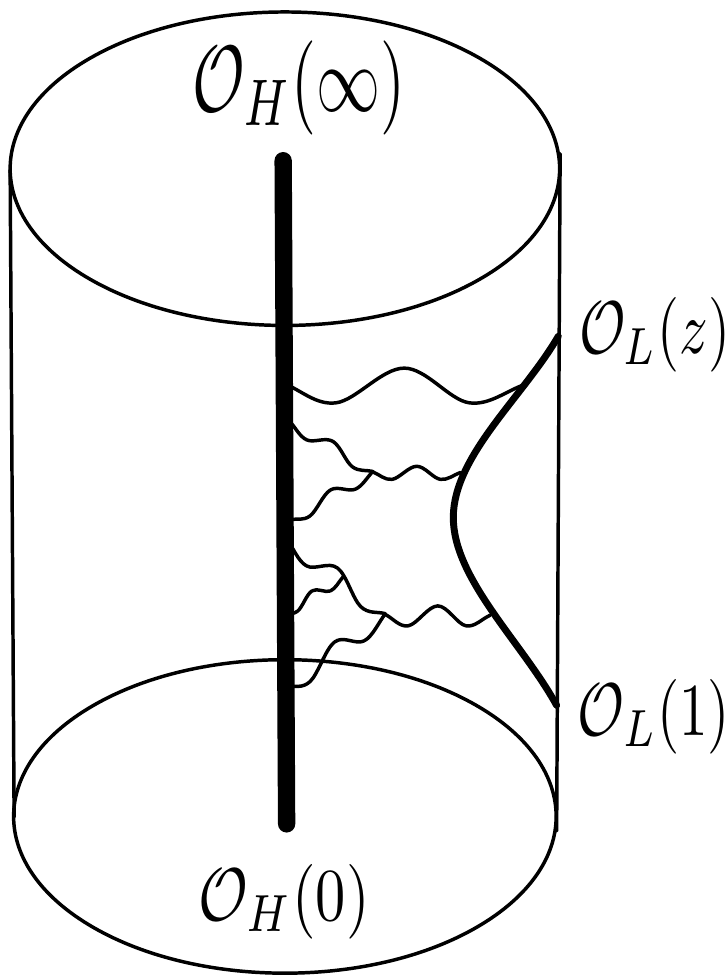}
\caption{ This figure provides a suggestive depiction of how graviton exchanges in AdS build up the classical field experienced by a light probe. We will compute the `graviton' exchanges in the CFT$_2$ by explicitly summing over the exchange of all multi-stress tensor operators. Note that our computation is not equivalent to the calculation of a sum of Witten diagrams; decomposed in terms of the exchange of states, Witten diagrams include both graviton and double-trace operator exchanges (as first noted in \cite{Liu}, and later substantially developed and systematized in \cite{DO1,DO2}).}
 \label{fig:IllustrationGravityDiagramsAdS} 
\end{center}
\end{figure}

In the AdS/CFT correspondence, the CFT stress tensor creates states that are dual to AdS gravitons.  Thus 
we say that an operator like $L_{-a_1} \cdots L_{-a_k}$ acts on the vacuum to create a $k$-graviton state.  When we sum over the exchange of all such (orthonormalized) states between a pair of local primary operators 
\be
\label{eq:SchematicVacuumBlock}
\mathcal{V}(z) = \left\< \CO_H(\infty) \CO_H(1) \left( \sum_{\{a_k \},\{ b_l\}}  \frac{  L_{-a_1} \cdots L_{-a_k} | 0 \>  \< 0 | L_{b_l} \cdots L_{b_1}  }{\mathcal{N}_{\{b_l \},\{ a_k\}}} \right) \CO_L(z) \CO_L(0) \right\>,
\ee
we obtain the Virasoro vacuum conformal block, which we define precisely in section \ref{sec:DiagramsandSummation}.   In this note we compute the Virasoro vacuum block by performing this sum explicitly.  We work in a heavy-light, large central charge limit where $c \to \infty$ with $h_H / c$ and $h_L$ fixed.   In AdS this would be interpreted as the exchange of all graviton states between a heavy object and a light probe, as seen in figure \ref{fig:IllustrationGravityDiagramsAdS}.

The strategy is as follows.  First we determine a basis of sufficiently orthonormalized $k$-graviton operators.  Then we show that when these operators act as in equation (\ref{eq:SchematicVacuumBlock}), the result can be reorganized into a set of tree diagrams with simple propagators and trivalent vertices; an example diagram is pictured in figure \ref{fig:VirasoroDiagramExample}.  Finally, we show that the sum over all diagrams obeys the differential recursion relation
\be
\label{eq:IntroRecursion}
\frac{1}{2 h_L} \left( \frac{d}{dz} \right)^2 \log \mathcal{V}(z) =  \frac{\epsilon}{(1-z)^2} +  \frac{1}{4 h_L^2} \left( \frac{d}{dz} \log \mathcal{V}(z) \right)^2,
\ee
where $\epsilon \equiv \frac{6 h_H}{c}$. 
There is an immediate connection with the Catalan numbers if we make the ansatz $\frac{1}{2h_L}\log {\cal V}(z) \stackrel{z \sim 1}{\sim} -B(\epsilon) \log (1-z)$, which becomes a good approximation for $z \to 1$.  Then equation (\ref{eq:IntroRecursion})  implies
\be
B(\epsilon) &=& \left[ B(\epsilon) \right]^2 + \epsilon, 
\ee
which is the recursion relation for the generating function $B(\epsilon)$ of the Catalan numbers $C_k$, where $C_k$ appears as the coefficient of $\epsilon^{k+1}$.
 Written in terms of $t \equiv \log(1-z)$, the full solution to equation (\ref{eq:IntroRecursion})  is the vacuum block
\be
\mathcal{V}(t) = e^{h_L t}  \left( \frac{\pi T_H}{\sin(\pi T_H t)} \right)^{2h_L},
 \ee
where 
\be
 T_H = \frac{1-2 B(\epsilon)}{2 \pi i} = \frac{\sqrt{\frac{24 h_H}{c} - 1}}{2 \pi}
\ee
determines the periodicity\footnote{The overall factor of $e^{h_L t}$ arises because we are studying the CFT in radial quantization on the Euclidean plane; it would be absent if we mapped the CFT to the cylinder, where equal time circles have constant circumference.} in Euclidean time $t$.

Thus the Hawking temperature of the dual BTZ black hole arises from evaluating the generating function of the Catalan numbers at $\frac{6 h_H}{c}$ and analytically continuing to $h_H > \frac{c}{24}$.  The Virasoro vacuum conformal block dominates correlation functions in the lightcone OPE limit \cite{Fitzpatrick:2012yx, KomargodskiZhiboedov, Fitzpatrick:2014vua}, so if it's appropriate to associate a temperature to the heavy state, then $T_H$ must be that temperature.  The computation of the vacuum block provides a kind of alternative derivation of the Cardy formula, which can be obtained through the solution of the thermodynamic relation $T = \frac{d E}{dS}$ if the microcanonical and canonical ensembles coincide.  

The diagrammatic rules arise out of the structure of the Virasoro algebra in a natural physical limit, after some very non-trivial simplifications and rearrangements discussed in section \ref{sec:DiagrammaticRules}.    Although the heavy-light Virasoro blocks were previously obtained by some of us \cite{Fitzpatrick:2014vua} in greater generality \cite{Fitzpatrick:2015zha}, our methods in this note are very different.  Aside from further insight into the structure of CFT$_2$, we hope that one of our methods can shed light on thermodynamics in higher dimensional CFTs and their relationship with black holes.  We give a brief discussion in section \ref{sec:CommentsHighTandD}, with more to come in future work.

This note provides an example where `on-shell methods' are simpler than Feynman diagrams,\footnote{In the context of scattering in flat spacetime, on-shell methods only utilize physical particle states associated with the `holographic' theory at null infinity.  In this sense, CFT states and methods provide on-shell computations of AdS observables.  More specifically, the $L_{n}$ create on-shell AdS$_3$ gravitons.} and where a perturbative diagrammatic expansion can be summed to all orders to reveal non-perturbative information about a non-linear classical field.  Although perturbative post-Newtonian type approximations have been well-developed \cite{Duff:1973zz}, this may be the only circumstance where such a resummation can capture all features of the exact classical solution for a field theory with non-linear interactions.

In  section  \ref{sec:DiagramsandSummation} we briefly review the Virasoro algebra and its conformal blocks.  Then we derive our diagrammatic rules and the associated differential recursion relation.  Along the way we also discuss some simplifying limits.  In section \ref{sec:AdS3Interpretation} we discuss the AdS$_3$ interpretation and the relation between our on-shell diagrammatic rules and bulk perturbation theory.  We discuss the results and future directions in section \ref{sec:Discussion}.  We relegate many of the technical details of the diagrammatic rules' derivation to appendix \ref{sec:AppendixDetailsDerivationRules}.

\section{Perturbative On-Shell Diagrams and Their All-Orders Summation}
\label{sec:DiagramsandSummation}

In any quantum mechanical system, we can insert $\bf 1$ as a sum over states and rewrite a correlation function as
\be
\< \CO_H(\infty) \CO_H(1) \left( \sum_{\psi} | \psi \> \< \psi | \right) \CO_L(z) \CO_L(0) \>.
\label{eq:GeneralSumStates}
\ee
Organizing this sum into irreducible representations of the theory's spacetime symmetry group defines a partial wave decomposition.  The partial waves themselves are purely kinematical functions that follow from symmetry, but their coefficients are dynamical data that characterize the theory.  In CFTs this is the conformal block decomposition, and the coefficients of each conformal block are products of operator product expansion (OPE) coefficients.  Two-dimensional CFTs have an infinite dimensional Virasoro algebra of symmetries, and so correlators can be written as a sum of Virasoro conformal blocks, weighted by the OPE coefficients of Virasoro primary (highest weight) operators.  

The vacuum state makes a universal contribution to equation (\ref{eq:GeneralSumStates}).  The global conformal generators $L_{0}$ and $L_{\pm 1}$, which correspond with dilatations, translations, and special conformal transformations, annihilate the CFT$_2$ vacuum. However, the vacuum `dynamically breaks' the other Virasoro symmetries, so that $L_{-n}$ with $n \geq 2$ act on the vacuum to create physical states.  These states are created by the stress energy tensor $T(z)$ and its derivatives, so we will use AdS/CFT inspired language and refer to them as `gravitons', since the stress tensor of the CFT is dual to the gravitational field in AdS.  We will be summing over all $k$-graviton states $| \psi \>$.  These are Virasoro descendants of the vacuum, and their sum is the Virasoro vacuum conformal block.  We say `Virasoro conformal block' to distinguish it from the global conformal blocks, which are based on only the global conformal symmetries $L_{\pm 1}$ and $L_0$. 

We are specifically studying CFT$_2$ at large central charge $c \gg 1$.  We include an unconventional factor of $\frac{1}{\sqrt{c}}$ in the normalization of the Virasoro generators, so that they satisfy
\be
\label{eq:VirasoroAlgebra}
[ L_n, L_m ] = \frac{n(n^2-1) }{12} \delta_{n,-m} + \frac{1}{\sqrt{c}}(n-m) L_{n+m}.
\ee
At very large $c$ the last term is suppressed, and so the Virasoro generators $L_{\mp n}$ approximate independent creation and annihilation operators.  We therefore obtain a basis for all Virasoro descendants of the vacuum, 
\be
| \{ a_k \} \> \equiv L_{-a_1} \cdots L_{-a_k} | 0 \>.
\ee
At infinite $c$ the $L_{-a_i}$ commute, so we take $2 \leq a_1 \leq a_2 \leq \cdots \leq a_k$ to avoid double-counting the states, although later on we will instead sum over all $a_i$ and divide by appropriate combinatorial factors.  At finite $c$ these states continue to provide a complete basis, but they are not orthonormal.  We denote their matrix of inner products as
\be
\mathcal{N}_{ \{b_l\}, \{ a_k \} } \equiv \< L_{b_l} \cdots L_{b_1} L_{-a_1} \cdots L_{-a_k} \>,
\ee
where we recall that $L_{n}^\dag = L_{-n}$ in radial quantization.  Our goal is to compute the Virasoro vacuum block, which we define precisely as
\be
\label{eq:ExactVacuumBlock}
\mathcal{V}(z) \equiv \< \CO_H(\infty) \CO_H(1)  \sum_{\{ a_i \}, \{b_j\}}  \left( | \{ a_k \} \> \mathcal{N}_{ \{b_l\}, \{ a_k \} } ^{-1} \< \{ b_l \} |  \right) \CO_L(z) \CO_L(0) \>
\ee
in the heavy-light limit $c \gg 1$, with $h_H \propto c$ and $h_L \ll c$.

\subsection{Approximating the Vacuum Block in the Heavy-Light Limit}
\label{sec:ApproximatingV}

Our main task is to simplify and then calculate the Virasoro vacuum block of equation (\ref{eq:ExactVacuumBlock}).  Let us first consider the ingredients that enter into the computation; for a more extensive review see e.g. appendix B of \cite{Fitzpatrick:2014vua}.

The matrix elements of $L_{-m}$ with $\CO_L$ and $\CO_H$ follow directly from the Ward identities for the stress tensor $T(z)$.  For a general primary operator $\CO(z)$ with holomorphic dimension $h$, we have
\be
\label{eq:LonO}
[ L_{m}, \CO(z) ] = 
z^m \left( \frac{ h (1+m)   + z \partial_z  }{\sqrt{c}} \right)  \CO(z).
\ee
It is straightforward to use this relation to compute matrix elements like $\<  L_{b_l} \cdots L_{b_1}   \CO_L \CO_L \>$ by commuting the $L_b$ through the $\CO_L$ until they annihilate the vacuum.  The normalizations $\mathcal{N}_{ \{b_j\}, \{ a_i \} }$ can be computed from the Virasoro algebra itself, equation (\ref{eq:VirasoroAlgebra}).

In the heavy-light limit the parameters $h_H \propto c \gg 1$, so we would like to first focus on where these parameters appear in the computation of $\mathcal{V}(z)$.  The action of $L_{-m}$ on $\CO_H(\infty) \CO_H(1)$ produces factors of $h_H$, so that
\be
\label{eq:HeavyMatrixElement}
\< \CO_H(\infty) \CO_H(1) | \{ a_k \} \> = \left( \frac{h_H}{\sqrt{c}} \right)^k \prod_{i=1}^k (a_i - 1) \ + \  O\left(h_H^{k-1}\right).
\ee
In other words, this matrix element is a polynomial of degree $k$ in $h_H$.  We have only displayed the leading term, as we will see that the other terms can be neglected in the heavy-light limit.  

Now let us examine the normalization matrix in more detail.  Since $L_0$ is a conserved charge, the matrix $\mathcal{N}_{ \{b_l\}, \{ a_k \} }$ vanishes unless the sums of the $a_i$ and $b_j$ are equal.  The diagonal part of $\mathcal{N}$ is
\be
\label{eq:kStateNorm}
\mathcal{N}_{ \{a_k\}, \{ a_k \} } = 
\prod_{i=1}^k \frac{a_i \left( a_i^2-1 \right)}{12},
\ee
where for simplicity we have assumed that the $a_i$ are distinct.    All non-diagonal matrix elements will be suppressed by factors of $1/\sqrt{c}$.  The reason is that we calculate
\be
\< 0 | L_{b_l} \cdots L_{b_1}  L_{-a_1} \cdots L_{-a_k} | 0 \>
\ee
by commuting the $L_{b_j}$ through the $L_{a_i}$ until they hit the vacuum ket and annihilate it.  If $a_i \neq b_i$ for any $i$, then we must use the final term in the Virasoro commutation relations of equation (\ref{eq:VirasoroAlgebra}) at least once, giving at least one factor of $1/\sqrt{c}$.  We also see that if the number of $a_i$ and $b_j$ differ, so $k \neq l$, then 
\be
\label{eq:NormParametrics}
\mathcal{N}_{ \{b_l\}, \{ a_k \} } \lesssim O \left( c^{-\frac{|k-l|}{2}} \right),
\ee
because we have to apply the last term in equation (\ref{eq:VirasoroAlgebra}) at least $|k - l|$ times.  

There is an additional simplification that is easiest to see by example.  Let us compare the `3-to-2' matrix elements
\be
\< L_{2} L_2 L_2 L_{-2} L_{-4} \>  = \frac{9}{2 \sqrt{c} }
\ee
versus
\be
 \< L_2 L_2 L_2 L_{-3} L_{-3} \> = \frac{30}{\sqrt{c^3} }. 
\ee
The second is negligible compared to the first in the large $c$ limit.
The distinction comes from the fact that we can extract $\< L_2 L_{-2} \>$ immediately from the first matrix element, and then we only need to apply the last term in the Virasoro commutation relations in equation (\ref{eq:VirasoroAlgebra}) once.  However, in the second matrix element we need to apply the last term from equation (\ref{eq:VirasoroAlgebra}) three times to rearrange the labels and obtain a non-vanishing matrix element.   

In general, any matrix element where the labels $b_i$ on the $L_{b_i}$ cannot be combined into groups and summed to exactly match the labels $a_j$, or vice versa, will be suppressed by additional factors of $\frac{1}{\sqrt{c}}$ beyond the minimum of $|k-l|$. These additional factors follow because we must apply the suppressed term in the Virasoro commutation relations more than $|k - l|$ times to reorganize the labels on the $L_{-n}$ into matching $L_{m} L_{-m}$ pairs.  We will see that only the leading terms in $\mathcal{N}_{ \{b_l\}, \{ a_k \} }$, namely those of of order $\frac{1}{\sqrt{c}^{|k-l |}}$, will contribute in the heavy-light limit. 

Now consider the implications of these results for computing the Virasoro vacuum block using equation (\ref{eq:ExactVacuumBlock}).  The matrix elements
\be
\< \CO_H(\infty) \CO_H(1)   |  \{ a_k \} \>  \propto  \left( \frac{h_H}{\sqrt{c} } \right)^k  \ \ \ \mathrm{and}  \ \ \ 
\< \{ b_l \} |   \CO_L(z) \CO_L(0) \> \propto \left( \frac{1}{\sqrt{c} } \right)^l
\ee
at large $h_H$ and $c$, with $h_H/c$ and $h_L$ fixed.  This combined with equation (\ref{eq:NormParametrics}) immediately implies that $l \leq k$, because otherwise the contribution to $\mathcal{V}(z)$ would vanish in the heavy-light limit.  It also means that we can ignore sub-leading terms in equation (\ref{eq:HeavyMatrixElement}).   However, we definitely do need to include $1/c$ suppressed effects in $\mathcal{N}_{ \{b_l\}, \{ a_k \} }$ when $k > l$.

In other words, we need to orthonormalize the $L_{b_i}$ that act on the light operators, keeping only effects of order $c^{-\frac{|k - l |}{2}}$.  This can be straightforwardly accomplished via Gram-Schmidt orthogonalization.  As a basis for $l$-graviton states, we can recursively define operators
\be
\label{eq:DefX}
X_{\{  b_l \}}  
= X_{\{ b_{l-1} \}} L_{b_l}  - \sum_{j=1}^{l-1} \frac{\< L_{b_j} L_{b_l} L_{-b_l-b_j} \>}{\< L_{b_l+b_j} L_{-b_l-b_j}\>} X_{\{ b_{l-1}, \hat b_j, b_l+b_j \}},
\ee
where $\{ b_{l-1}, \hat b_j, b_l+b_j \}$ denotes the set $b_{l-1}$ with the element $b_j$ replaced with $b_l+b_j$. As an example, in the 2-graviton case this is just
\be
X_{b_1, b_2} = L_{b_1} L_{b_2} - \frac{\< L_{b_1} L_{b_2} L_{-b_1 - b_2}  \> }{\< L_{b_1+b_2} L_{-b_1 - b_2}  \> } L_{b_1 + b_2},
\ee
which can easily be seen to be orthogonal to $L_{-b_1 - b_2}$.  This operator is also sufficiently orthogonal to all other 1 and 2 graviton states.  The recursive definition implies that all of the $X_{\{b_l \} }$ are sufficiently orthogonal at leading order in the heavy-light limit.

Using these new orthogonalized operators,  we can rewrite the Virasoro vacuum block in the heavy-light limit as
\bq
\label{eq:ApproximateVacuumBlock}
\begin{split}
\mathcal{V}(z) &\approx \< \CO_H(\infty) \CO_H(1)  \sum_{\{ a_i \}}  \left( \frac{ | X_{\{ a_k \}} \>\<  X_{\{a_k \} }| }{  \mathcal{N}_{ \{a_k\}, \{ a_k \} } }  \right) \CO_L(z) \CO_L(0) \> \\
&\approx \< \CO_H(\infty) \CO_H(1)  \sum_{\{ a_i \}}  \left( \frac{ | \{ a_k \} \>\<  X_{\{a_k \} }| }{  \mathcal{N}_{ \{a_k\}, \{ a_k \} } }  \right) \CO_L(z) \CO_L(0) \>.
\end{split}
\eq
The challenge that remains is to evaluate the matrix elements of the $X_{\{a_k \} }$ operators.  Although the individual matrix elements are very complicated, they can be rewritten in terms of a set of simple diagrammatic rules, as we discuss in the next section.

Before moving on to the diagrammatics, we note one additional simplification.  As we already proved by very different methods \cite{Fitzpatrick:2015zha}, in the heavy-light limit we can write the vacuum Virasoro conformal block as
\be
\label{eq:Exponentiation}
\mathcal{V}(z) = \exp \left[ 2 h_L f \left( \frac{h_H}{c}, z \right) \right],
\ee
for some function $f$ that we will determine.  This means that we only need to study the terms in equation (\ref{eq:ApproximateVacuumBlock}) that are proportional to a single power of $h_L$; in what follows we will focus on precisely these terms.  Later on we will explain how this exponentiation can be understood using the methods of this paper.

\subsection{Diagrammatic Rules for the Vacuum Block}
\label{sec:DiagrammaticRules}

\begin{figure}[t!]
\begin{center}
\includegraphics[width=0.95\textwidth]{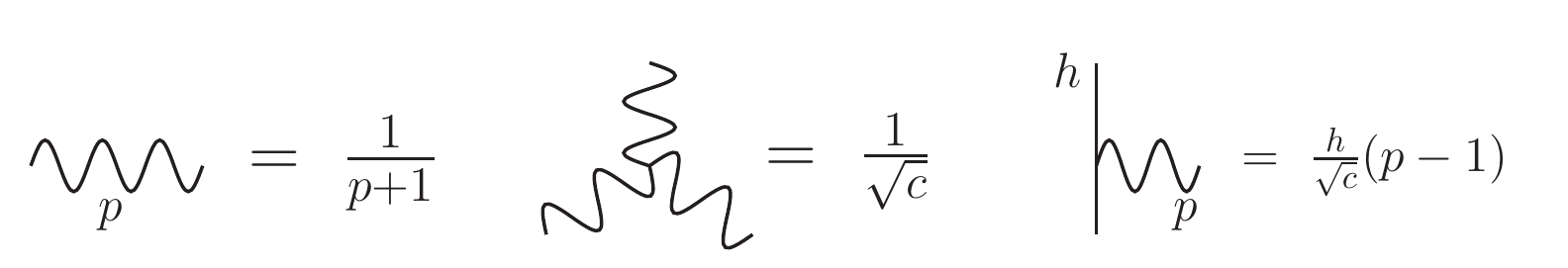}
\caption{ This figure summarizes the diagrammatic rules used to construct the Virasoro vacuum block. The semi-classical Virasoro block for $\<\Ocal_H \Ocal_H \Ocal_L \Ocal_L\>$ is constructed by exponentiating all diagrams with an arbitrary number of initial gravitons coupled to the heavy operator, which cascade into one final graviton that couples to the light operator. An example of one such diagram is shown in figure \ref{fig:VirasoroDiagramExample}.}
\label{fig:VirasoroRules} 
\end{center}
\end{figure}

\begin{figure}[t!]
\begin{center}
\includegraphics[width=0.60\textwidth]{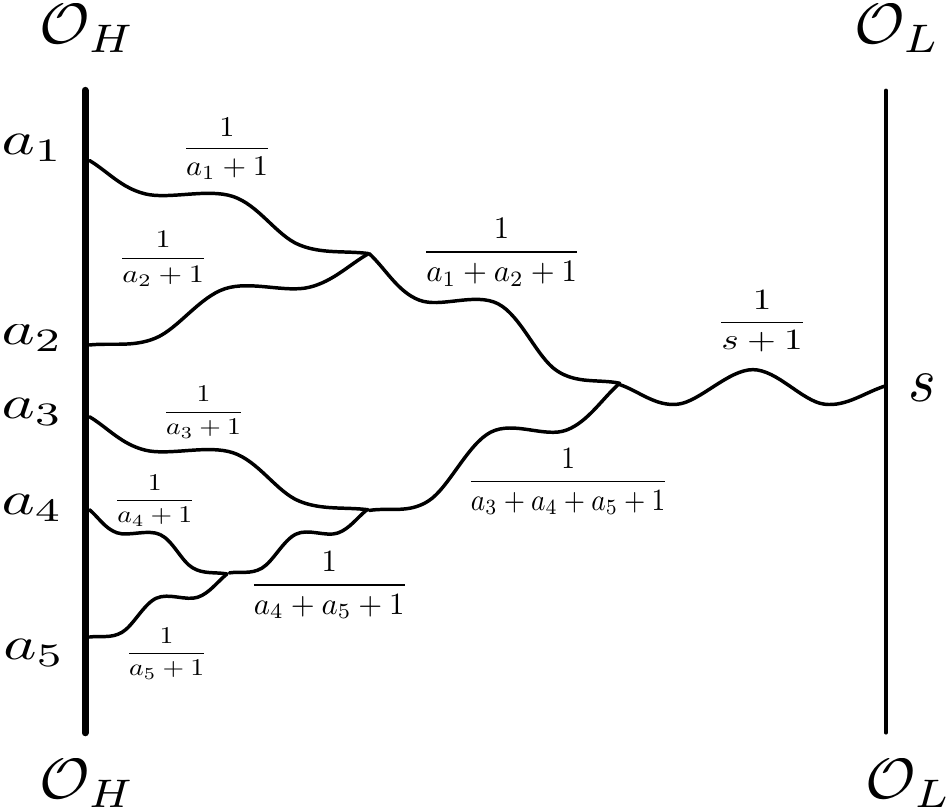}
\caption{ This figure indicates how the Virasoro diagrammatic rules work for a 5-to-1 diagram. To complete the rules, we multiply all the displayed propagators by the vertex rules displayed in figure \ref{fig:VirasoroRules} and an overall factor of $2 \left( \frac{6^5 z^s}{s(s-1)} \right)$, where $s$ is the sum of all $a_i$. }
 \label{fig:VirasoroDiagramExample} 
\end{center}
\end{figure}

To obtain a single term in the sum in our approximate expression for the vacuum block, equation (\ref{eq:ApproximateVacuumBlock}), we need only follow some simple diagrammatic rules, which are depicted schematically in figure \ref{fig:VirasoroRules}.  All diagrams are trees with trivalent vertices.  Figure \ref{fig:VirasoroDiagramExample}  shows an example of one such diagram, labeled with its propagators.  We will only study diagrams that end with a single propagator connecting to the light operators, since all other diagrams can be obtained via exponentiation, as indicated in equation (\ref{eq:Exponentiation}).  We will first state the diagrammatic rules, and then outline their derivation, leaving many details to appendix \ref{sec:AppendixDetailsDerivationRules}:  
\begin{enumerate}
\item  Label the $k$ initial gravitons connected to the heavy operators with integers $a_1, \cdots, a_k$.  
\item Draw all binary tree diagrams where the $k$ initial gravitons combine via 3-pt vertices to become a single graviton, which connects with the light operators. 
\item For each `propagator' define its momentum $p$ as the sum of the $a_i$ flowing through it; momentum is conserved at vertices.  Include a factor of
\begin{equation*}
\frac{1}{p+1}
\end{equation*}
for each propagator.  If the leg is initial then  $p=a_i$; figure \ref{fig:VirasoroDiagramExample} provides an example.
\item For each vertex coupling a graviton of momentum $p$ to an external operator with scaling dimension $h$, include a factor of
\begin{equation*}
\fr{h}{\sqrt{c}} (p-1)
\end{equation*}
while for each 3-graviton vertex, include a factor of $\fr{1}{\sqrt{c}}$.
\item Take the product of the propagators and then multiply the result by\footnote{A careful reader might have expected this prefactor to instead be proportional to $12^k$, due to the graviton normalization in eq.\ (\ref{eq:kStateNorm}). However, this normalization is then divided by a symmetry factor of $2^{k-1}$ due to the equivalence of swapping the ordering of any pair of tree branches that meet at any given vertex, leaving the overall factor of 2 consistent with eq.\ (\ref{eq:Exponentiation}).}
\be
\label{eq:OverallFactorinRules}
2\cdot \frac{6^k z^s}{s(s-1)},
\ee
where $s \equiv \sum_i a_i$.
\item Sum the resulting tree diagrams over all $a_i$ from $2$ to $\infty$ to obtain the $k$-to-$1$ contribution to the Virasoro vacuum block.
\end{enumerate}

Before discussing the proof, let us consider the case of $k=1,2$ gravitons as elementary examples.  The $1$-graviton sector of equation (\ref{eq:ApproximateVacuumBlock}) simply gives
\be
\sum_{a=2}^\infty \frac{12 h_H h_L}{c} \frac{(a-1)(a-1)}{a(a^2-1)} z^a = 2 \sum_{a=2}^\infty  \frac{6 z^a}{a(a-1)} \fr{h_H}{\sqrt{c}} (a-1)  \frac{1}{a+1} \fr{h_L}{\sqrt{c}} (a-1) .
\ee
We have grouped the terms to emphasize that this accords with the diagrammatic rules at $k=1$.  The two factors of $(a-1)$ in the numerator of the left-hand side come from connections to the heavy and light operators, while the denominator comes from the normalization of the 1-graviton state. The sum can be performed to give the global conformal block for stress tensor exchange in a CFT$_2$, which is proportional to $z^2 {}_2F_1(2,2,4,z)$.  

To study the $k=2$ case, we need to compute and simplify the $\<X_{a_1, a_2} \CO_L \CO_L \>$ correlator.  We can evaluate the Virasoro matrix element in the definition of $X_{a_1, a_2}$ to give
\be
X_{a_1, a_2} = L_{a_1} L_{a_2} -  \fr{1}{\sqrt{c}} \frac{(a_1 + 2 a_2)a_1(a_1^2-1)}{(a_1+a_2)((a_1+a_2)^2-1)} L_{a_1 + a_2}.
\ee
We are only interested in the term proportional to $h_L$ in the correlator with the light operators; using equation (\ref{eq:LonO}) we have
\bq
\label{eq:2GravitonSimplification}
\begin{split}
\<&X_{a_1, a_2} \CO_L(z) \CO_L(0) \> \\
&= \left( \fr{h_L}{c} a_1 (a_1-1) - \fr{1}{\sqrt{c}} \frac{(a_1 + 2 a_2)a_1(a_1^2-1)}{(a_1+a_2)((a_1+a_2)^2-1)} \fr{h_L}{\sqrt{c}} (a_1 + a_2 - 1) \right) z^{a_1+a_2} + {\cal O}(h_L^2) \\
&= \fr{h_L}{c} \frac{a_1 (a_1 - 1)}{s(s+1)} \Big( s(s+1)  - (s + (a_2 - 1)) (s - (a_2 -1))   \Big) z^s + {\cal O}(h_L^2) \\
&= \fr{h_L}{c} \frac{a_1 (a_1 - 1) a_2 (a_2-1)}{s(s+1)} z^s + {\cal O}(h_L^2)
\end{split}
\eq
for the light matrix element.
These factors combine with the heavy matrix element and the graviton normalizations to yield the result implied by the diagrammatic rules for the single 2-to-1 binary tree diagram.  The order $h_L^2$ terms just give the square of the 1-graviton matrix element, in keeping with the exponentiation in equation (\ref{eq:Exponentiation}).

Now let us discuss the general proof for $k$ gravitons.  The idea is to prove the rules inductively by assuming that matrix elements of $X_{k-1}$ can be computed from the diagrammatic rules, and then to use equation (\ref{eq:DefX}) to show that matrix elements of $X_k$ will also be obtained from the rules.  We already saw this occur when we simplified the 2-graviton case in equation (\ref{eq:2GravitonSimplification}), but the general case is more involved.  We will leave the detailed algebra to appendix \ref{sec:AppendixDetailsDerivationRules}, and only describe the process in words and pictures here.

\begin{figure}[t!]
\begin{center}
\includegraphics[width=0.95\textwidth]{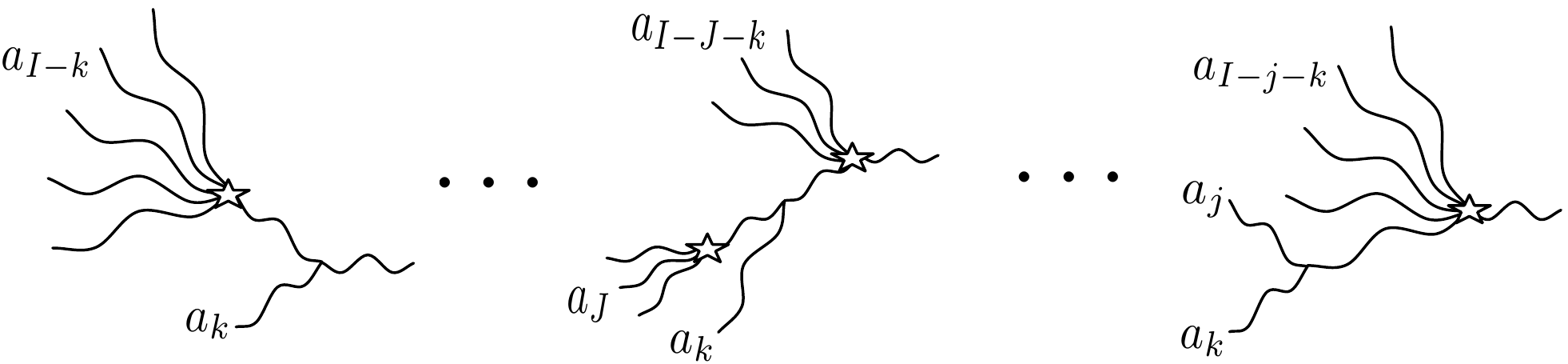}
\caption{ This figure indicates what kinds of structures can arise when we add a $k^{th}$ graviton to a $k-1$ graviton diagram. Stars indicate a sum over possible sub-diagrams.  The diagram on the left is of the `trunk' type, while the one on the right is the opposite extreme, where $k$ attaches to a single other graviton.  The central diagram lies between these two extremes.  In essence, the algebraic manipulations of appendix \ref{sec:AppendixDetailsDerivationRules} massage diagrams of the first and last types (which are accompanied by various prefactors) into a sum over diagrams of all types, redistributing the $k$th graviton into all possible configurations.}
 \label{fig:VirasoroDiagramStructures} 
\end{center}
\end{figure}

Consider adding a $k$th graviton to a $k-1$ graviton diagram.  Roughly speaking, there are three types of diagrams that can then occur, and they are pictured in figure \ref{fig:VirasoroDiagramStructures}, to which we will refer.  
The $k$th graviton can attach to the `trunk' of the $k-1$ graviton diagram, or it can attach somewhere in the middle, or it can attach to one of the `leaves'.  However, the terms in equation (\ref{eq:DefX}) only produce `trunk' and `leaf' attachments, and they come with algebraic prefactors. Specifically, the term $X_{\{ a_{k-1} \}} L_{b_k}$ produces a `trunk' attachment as on the far left, multiplied by a prefactor, while the sum over the $j$ terms $X_{\{ a_{k-1}, \hat a_j, a_k+a_j \}}$ take the form of the `leaf' attachments on the far right, again multiplied by prefactors.  The algebraic manipulations discussed in appendix \ref{sec:AppendixDetailsDerivationRules} make it possible to re-express these special classes of diagrams and their prefactors as a sum over all three types of $k$-to-1 diagrams.  This leads to an inductive proof of the diagrammatic rules for any $k$.

Finally, let us comment on the exponentiation of the diagrams encapsulated in equation (\ref{eq:Exponentiation}), which is best demonstrated through a simple example. Consider the 2-graviton contribution before taking the heavy-light limit $h_H \gg h_L$,
\be
&&\<\Ocal_H \Ocal_H \left( \fr{|X_{a_1,a_2}\>\<X_{a_1,a_2}|}{\Ncal_{a_1,a_2}} \right) \Ocal_L \Ocal_L\> = \left( \fr{12 h_H h_L}{c} \fr{(a_1-1)z^{a_1}}{a_1(a_1+1)} \right) \left( \fr{12 h_H h_L}{c} \fr{(a_2-1)z^{a_2}}{a_2(a_2+1)} \right) \qquad \, \, \\
&& \, + \, \fr{144 h_H^2 h_L}{c^2} \fr{(a_1-1)(a_2-1)z^s}{(a_1+1)(a_2+1)s(s+1)} + \fr{144 h_H h_L^2}{c^2} \fr{(a_1-1)(a_2-1)z^s}{(a_1+1)(a_2+1)s(s+1)} + O(h_H h_L). \nn
\ee
The first term in this expression is simply the product of two copies of 1-graviton exchange, which is consistent with exponentiation. The next term is the same 2-to-1 contribution computed in eq.\ (\ref{eq:2GravitonSimplification}).

The third term has almost the same structure as the second, but with $h_H \lra h_L$, indicating that it corresponds to a `1-to-2' diagram where one initial graviton emitted by $\Ocal_H$ splits into two final gravitons coupled to $\Ocal_L$. Though such a term can be constructed using our Virasoro rules, it vanishes as $1/c$ in the heavy-light limit, as does the remaining $O(h_H h_L)$ term. We therefore see that the only non-vanishing 2-graviton contributions are those that can be written in terms of diagrams which are linear in $h_L$.

More generally, consider taking any $k$-to-1 diagram and adding a second graviton which couples to $\Ocal_L$. We then have two choices for where to couple the other end of this new propagator: to one of the other exchanged gravitons or directly to $\Ocal_H$. Following our Virasoro rules, coupling to another graviton adds an overall factor of $\fr{h_L}{c}$, while coupling to the heavy operator changes this factor to $\fr{h_H h_L}{c}$. The \emph{only way} to couple a second graviton to the light operator (without the diagram vanishing) is therefore to start a \emph{new} tree of gravitons. However, given our rules, such a diagram can be written as simply the product of two single-tree diagrams.

Exponentiation in the heavy-light limit therefore follows directly from our Virasoro rules. Adding a new graviton to a diagram always `costs' a factor of $1/c$, but this suppression is compensated by a factor of $h_H$ whenever the graviton couples to $\Ocal_H$. The only non-vanishing diagrams are therefore those that can be written as products of asymmetric $k$-to-1 trees, each of which has an overall factor of $2h_L$. We then only need to exponentiate the set of single-tree diagrams to obtain the full Virasoro block. This exponentiation can also be derived using other methods \cite{Fitzpatrick:2014vua,Fitzpatrick:2015zha}.

\subsection{Summing Diagrams with a Recursion Relation}

We have seen that the Virasoro vacuum block can be computed using diagrammatic rules.  But this still leaves us with the problem of summing up the diagrams and the discrete $a_i$ `momenta' flowing through them.  This can be accomplished by thinking about $f = -\frac{1}{2 h_L} \log \mathcal{V}$ as a generating function for trees satisfying a recursion relation.

Let us first review the much simpler question of counting binary trees.  Let $B(\epsilon)$ be the generating function for the number of binary trees, so that the coefficient of $\epsilon^n$ in $B(\epsilon)$ is the number of trees with $n$ leaves (or equivalently, $n-1$ vertices):
\be
B(\epsilon) &\equiv& \sum_{n=0}^\infty C_n \epsilon^{n+1}.
\ee  
All binary trees with more than one leaf divide at their trunk into two branches, which are themselves binary trees.
This means that $B(\epsilon)$ satisfies the recursion relation
\be
B(\epsilon) =  B^2(\epsilon) + \epsilon,
\ee
where we must add $\epsilon$ to include the base case of a tree with only one leaf.   Fixing $B(0)=0$, this recursion relation is easily solved to give
\be
B(\epsilon) = \frac{1 - \sqrt{1 - 4 \epsilon} }{2},
\ee
which is the generating function of the Catalan numbers.

Now let us consider the $f(\epsilon, z)$ from equation (\ref{eq:Exponentiation}), where we write $\epsilon = \frac{6 h_H}{c}$ for convenience.  In the previous section we discussed how to compute $f$ using the diagrammatic rules.  Thus we can write
\be
\label{eq:UsefulFormofF}
f(\epsilon, z) = -\log (z) + \sum_{\mathrm{Diagrams}} ( \cdots \mathrm{rules} \cdots ) \epsilon^k \frac{ z^s}{s(s+1)},
\ee
where $s = \sum_i a_i$, and $k$ is the number of gravitons.  By a sum over `Diagrams' we imply both a sum over diagram topologies and over the $a_i$ `momenta'.   The term $-\log z$ accounts for the disconnected part of the heavy-light correlator, or in other words, the part that comes from the exchange of the vacuum itself between the heavy and light operators, which is just $z^{-2h_L}$.  

We have explicitly emphasized the $z$ and $s$ dependence in equation (\ref{eq:UsefulFormofF}) to make it clear that by differentiating $f$ with respect to $z$, we can remove the `trunk' from all tree diagrams.  Our claim is that $f$ satisfies a recursion relation analogous to that of the Catalan numbers,
\be
\label{eq:RecursionRelation}
\left( \frac{d}{dz} \right)^2 f(\epsilon, z) =  \frac{\epsilon}{(1-z)^2} + \left( \frac{d}{dz} f(\epsilon, z) \right)^2.
\ee
To see this, consider the derivatives of $f$, which we can write in schematic form as
\be
\left( \frac{d}{dz} \right) f(\epsilon, z) = -\frac{1}{z} + \sum_{\mathrm{Diagrams}} \left( \cdots \right) \epsilon^k \frac{z^{s-1}}{(s+1)} 
\ee
and
\be
\left( \frac{d}{dz} \right)^2 f_\epsilon(z) = \frac{1}{z^2} + \sum_{\mathrm{Diagrams}} \left( \cdots \right) \epsilon^k \frac{s-1}{s+1} z^{s-2}.
\ee
This last expression is the left-hand side of the claimed recursion relation, equation (\ref{eq:RecursionRelation}).  The right-hand side of the relation is
\be
\label{eq:RHSofRecursion}
\mathrm{RHS} =  \frac{ \epsilon}{(1-z)^2}  +  \frac{1}{z^2} - \sum_{\mathrm{Diagrams}} \left( \cdots \right) \epsilon^k \frac{2 z^{s-2}}{(s+1)} 
+ \left( \sum_{\mathrm{Diagrams}} \left( \cdots \right) \epsilon^k \frac{z^{s-1}}{(s+1)}  \right)^2.
\ee
Now we can move all but the last term from the RHS to the LHS.  
The $1/z^2$ terms cancel, as do the $k=1$ graviton contributions via
\be
\epsilon  \sum_{a=2} \frac{(a-1)^2}{a+1}   z^{a-2} - \frac{\epsilon}{(1-z)^2} + \epsilon  \sum_{a=2} \frac{2 (a-1)}{a+1} z^{a-2} = 0.
\ee
The other terms combine to become
\be
LHS \to \sum_{\mathrm{Diagrams}} \left( \cdots \right)  \epsilon^k \frac{(s-1) + 2}{s+1} z^{s-2} =  \sum_{\mathrm{Diagrams}} \left( \cdots \right)  \epsilon^k z^{s-2}.
\ee
Thus on the left-hand side we have the sum over all diagrams with the 1-graviton piece deleted, and with the factors of $\frac{1}{s(s+1)}$ associated with the `trunk' of the diagram removed.  This sum is therefore equal to the square of the sum of diagrams with only propagator factors in equation (\ref{eq:RHSofRecursion}).  So the differential recursion relation, equation (\ref{eq:RecursionRelation}), has been proved.

The differential recursion relation of equation (\ref{eq:RecursionRelation}) can be solved, yielding
\be
f(\epsilon, z) & = & -\left( \frac{1-\sqrt{1-4 \epsilon } }{2} \right) \log (1-z) - \log \left(\frac{1-(1-z)^{\sqrt{1-4 \epsilon }}}{\sqrt{1-4
   \epsilon }}\right)
\ee
which we previously obtained by other methods \cite{Fitzpatrick:2014vua, Fitzpatrick:2015zha}.  The generating function for the Catalan numbers makes a prominent appearance as the coefficient of $\log(1-z)$.  In fact, by keeping only the leading effects at large momenta $a_i$, one can show that the leading logarithmic behavior of $f$ as $z \to 1$ is the same for every diagram, so that the leading logarithm simply gets multiplied by the generating function of the number of diagrams as a function of $\epsilon$.

The result for $f$ may appear complicated, but if we define
\be
T_H \equiv \frac{\sqrt{4 \epsilon - 1}}{2 \pi},
\ee
for $\epsilon > 1/4$, and also use $t = \log(1-z)$, then we can exponentiate $f$ to yield
\be
\mathcal{V}(t) = e^{h_L t}  \left( \frac{\pi T_H}{\sin(\pi T_H t)} \right)^{2h_L}, 
\label{eq:Vformula}
\ee
as claimed in the introduction.  For $\epsilon < 1/4$ we obtain the analytic continuation of this result, where the sine becomes a sinh.  In the dual AdS$_3$ theory, this range of $\epsilon$ corresponds to a deficit-angle geometry  \cite{Fitzpatrick:2014vua}. Note that the overall factor of $e^{h_L t}$ arises because we are working in radial quantization on the plane, where circles have expanding radius $e^t$.  If we mapped the CFT to the cylinder this factor would be absent, and the correlator would be exactly periodic in Euclidean time, signaling thermality at a temperature $T_H$.

\subsection{Comments on High Temperature and Higher Dimensions}
\label{sec:CommentsHighTandD}

\begin{figure}[t!]
\begin{center}
\includegraphics[width=0.75\textwidth]{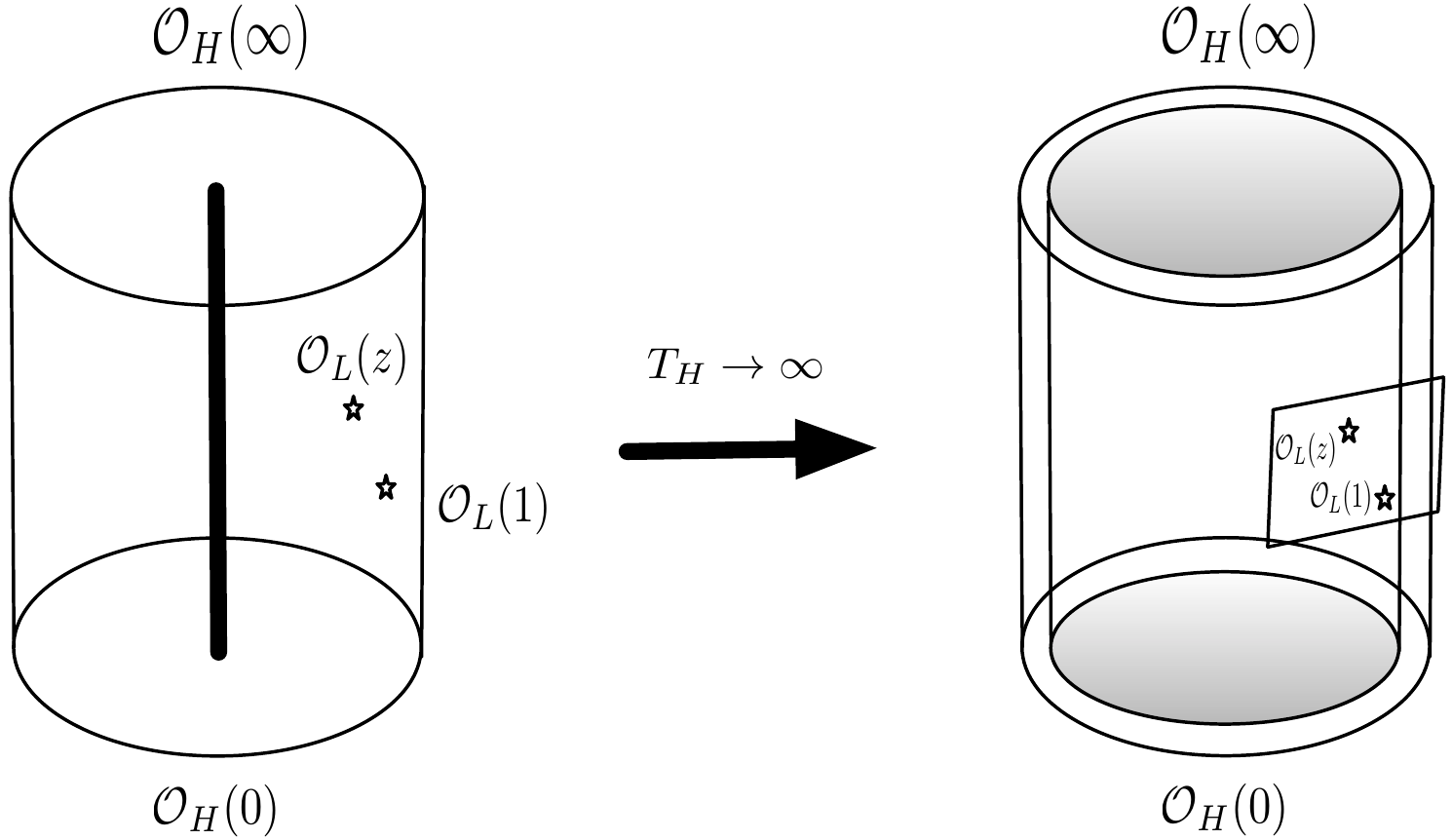}
\caption{ This figure depicts the high-temperature limit of a large AdS$_{d+1}$ black hole.   By studying short-distance physics in the CFT, with distances measured in units of the inverse temperature, we can recover the physics of a CFT in $R^d$ at finite temperature, which is dual to a flat black brane in Poincar\'e patch coordinates.}
 \label{fig:DiagramHighTemperatureLimit} 
\end{center}
\end{figure}

We have worked with $h_H / c$ as a fixed parameter, but it is interesting to study the limit $h_H / c \to \infty$, or in other words, the limit $T_H \to \infty$.  This limit has a simple interpretation in both the CFT and the putative dual AdS$_3$ theory, as indicated in figure \ref{fig:DiagramHighTemperatureLimit}.

In the high-temperature limit, it is natural to measure distances in the CFT in units of $T_H^{-1}$.  Performing this rescaling on the vacuum block by defining
\be
x \equiv  \pi T_H t = \pi  T_H \log(1-z).
\ee
Since $e^{-t h_L} {\cal V}(t)$ in (\ref{eq:Vformula}) just depends on $t$ and $T_H$ through the combination $x$, its {\it complete functional dependence} is preserved in this high temperature limit.  

It is interesting to consider which `gravitons' contribute in the high-temperature limit.  When we view $\mathcal{V}(z)$ as a series in small $z$, our rescaling sends $z \sim -\frac{x}{\pi T_H} + \cdots$ where the ellipsis denotes terms suppressed by more powers of $T_H^{-1}$.  In the high-temperature limit, these factors can only be cancelled by factors of $\frac{h_H}{c} \propto T_H^2$.  This means that only the term of order $z^{2k}$ survives when we study $k$-graviton exchange.  In other words, the only non-vanishing contributions are the exchange of the operators $L_{-2}^k$ and $X_{-2,\cdots, -2}$ as $T_H \to \infty$.  This means that it is the OPE coefficients
\be
\< \CO_H \CO_H T^k \> \ \ \ \mathrm{and}  \ \ \  \< \CO_L \CO_L T^k \>
\ee
that control the thermodynamic behavior of the correlator in the high-temperature limit at large $c$.  In particular, multi-stress tensor operators with derivatives will not contribute at large $T_H$.  In the case of CFT$_2$ these OPE coefficients are determined by the Virasoro algebra, but in higher dimensions they would be independent CFT data.  Thermodynamic expectations then suggest that these OPE coefficients may be universal at large central charge.  We will have more to say about the relationship between CFT data, thermodynamics, and the high temperature limit of general CFTs in future work.

\section{AdS$_3$ Interpretations}
\label{sec:AdS3Interpretation}

We discussed the importance of the Virasoro vacuum block in AdS$_3$/CFT$_2$ in previous work \cite{Fitzpatrick:2014vua,  Fitzpatrick:2015zha}.  In brief, the vacuum block can be used to measure the Hawking temperature of BTZ black holes \cite{Fitzpatrick:2014vua}, to measure the length of geodesics in asymptotically AdS$_3$ backgrounds \cite{Kraus:2002iv, Hijano:2015rla}, to determine the quasinormal mode spectrum in these backgrounds \cite{Fitzpatrick:2014vua}, to compute entanglement entropies \cite{Asplund:2014coa,Asplund:2015eha,Hartman:2013qma,Headrick,Headrick:2015gba}, and to compute Lyapunov exponents characterizing chaos \cite{Roberts:2014ifa}.  
In what follows we will consider an ad hoc approach where we guess a Lagrangian that produces the diagrammatic rules and then briefly discuss how the rules might be derived from AdS$_3$ gravity.

\subsection*{Guessing a Lagrangian\footnote{We thank Cliff Cheung for emphasizing this viewpoint and for discussions of this simple Lagrangian theory.}}

Instead of deriving the diagrammatic rules from AdS$_3$, we can try to simply guess a Lagrangian with an identical set of Feynman rules.  One might also be motivated to take this approach because in this case, the bulk theory truly `emerges' from the CFT$_2$ at large central charge in a very obvious way.  However, the theory that we will write down actually will only live in two dimensions.

We would like to find a theory with integer valued `momenta' and trivalent vertices.  So consider the action
\be
S = \int dt 
\int_0^{2 \pi} d \theta \left( \phi^\dag (\partial_t - \partial_ \theta  - 1) \phi + g \phi^\dag \phi(\phi^\dag + \phi) \right),
\ee
where we will work in perturbation theory in the coupling $g \propto \frac{1}{\sqrt {c}}$.  The field $\phi(t,\theta)$ has a first order equation of motion with a mode expansion
\be
\phi_m = e^{i (m+1) t - i m \theta}
\ee
labeled by integers $m$.  The frequencies and integer momenta are conserved at the trivalent vertices, and the theory has a propagator
\be
\frac{i}{ \omega - (m+1)}.
\ee
If we include the heavy and light operators as static external sources for the $\phi$ field, then the frequencies will all vanish \cite{Weinberg:1995mt} in the resulting correlator, but general integer momenta will be able to propagate, leading to a diagrammatic structure similar to that in section \ref{sec:DiagrammaticRules}.  It would be interesting to connect this ad hoc theory to either the bulk gravitational theory (presumably in the Chern-Simons description), or perhaps to a limit of Liouville theory.

\subsection*{Diagrammatic Rules from AdS$_3$ Gravity?}

Our diagrammatic rules are purely holomorphic, and involve only 3-pt interaction vertices.  Perturbation theory in fluctuations of the metric will not have either of these properties.  The well-known Chern-Simons description of gravity \cite{Witten:1988hc} represents the metric in terms of connections $A$ and $\bar A$ that can be purely holomorphic and anti-holomorphic, respectively, so it is a natural starting point.  However, even using the $sl(2)$ Chern-Simons representation of AdS$_3$ gravity, interactions between heavy matter sources and $A$ will be complicated.  The methods of \cite{Hijano:2015qja} provide a clever solution to this last problem -- one can study a higher spin theory in AdS$_3$, with both the `graviton' and the matter contained in the $sl(N)$ Chern-Simons field, with $N$ taken to be large.  In this case, all interactions will be through 3-pt vertices.

The general heavy-light Virasoro blocks have recently been obtained directly from AdS$_3$ \cite{Hijano:2015qja}.  The method works by studying bulk diagrams where the propagation of external states has been confined to geodesics \cite{Hijano:2015zsa}; this is justified as follows.  When the external states propagate on geodesics one computes (in effect) a correlator where the external operators have infinite dimension.  The conformal block decomposition of a Witten diagram involves the exchanged states plus double-trace operators made from the external states, but these decouple when those states have infinite dimension. So by evaluating geodesic Witten diagrams one computes the conformal block associated to the exchanged states.

Although this method can determine the blocks at leading order in $1/c$, it was based on classical solutions in AdS (deficit angles and BTZ black holes), so it does not explain the dramatic simplifications we have found in perturbation theory in $1/c$, and the associated diagrammatic rules.  Current conformal blocks have been obtained previously from Chern-Simons theories in AdS$_3$ \cite{Keranen:2014ava}, but again the computation was not based on perturbation theory. Thus far we have not found an elegant bulk derivation of the on-shell rules.    Much like on-shell recursion relations in flat spacetime \cite{Elvang:2013cua}, it is possible that the Virasoro rules cannot be straightforwardly derived from off-shell bulk diagrams, or perhaps a nice derivation will be found in the future.  The most likely route would be to use the higher-spin description \cite{Hijano:2015qja} and solve the flatness condition for the Chern-Simons fields order by order in $1/c$ perturbation theory.

\section{Discussion and Future Directions}
\label{sec:Discussion}

We have derived a set of diagrammatic rules that compute the vacuum Virasoro conformal block, and then summed the diagrams using a recursion relation.  The resulting vacuum block has a number of implications and interpretations relevant to eigenstate thermalization and AdS$_3$/CFT$_2$.  Let us discuss some possibilities for future work.

There are a large number of other theories and other kinematic limits in which similar diagrammatic rules may emerge:
\begin{itemize}
\item The most obvious generalization would be to general Virasoro conformal blocks in the heavy-light limit.  The results are known \cite{Fitzpatrick:2015zha} in closed form, which suggests that a diagrammatic derivation may be possible.
\item One might generalize our methods to theories where the Virasoro algebra is contained in a larger symmetry algebra.  There is already some work along these lines using other methods \cite{deBoer:2014sna}.  If the diagrammatic rules can be derived from the Chern-Simons description of AdS$_3$ gravity, then it is very likely that higher-spin theories in AdS$_3$ also have diagrammatic rules.
\item  It would be interesting to study the limit of large intermediate dimension with fixed $c$ and fixed external dimensions \cite{Maldacena:2015iua}.  Zamolodchikov showed  \cite{Zamolodchikovq} that in this limit the blocks take a very simple form in terms of the uniformizing variable $q$ (see \cite{HarlowLiouville} for a nice review), and it would be fascinating to see that variable emerge from diagrammatics.  
\end{itemize}

In section \ref{sec:DiagrammaticRules} we derived the rules by orthonormalizing the Virasoro generators at leading order in $1/c$.  However, that derivation required some algebraic gymnastics.  It would interesting to find a more streamlined argument, perhaps using a more natural basis of Virasoro generators.  This might also simplify the computation of corrections to the block in $1/c$ perturbation theory, and enable further generalizations to other kinematic limits, and to other theories.  It might also make it easier to connect with bulk perturbation theory.  At present we do not know how to construct the appropriate `loop diagrams' encoding $1/c$ corrections, but this might be possible after further study.

There are a host of older results on Virasoro conformal blocks, such as the Zamolodchikov recursion relations \cite{Perlmutter:2015iya} and the AGT relations \cite{Alday:2009aq}.  It would be fascinating to understand how our results connect with these other methods.  In particular, perhaps there is some connection between our recursion relation in equation (\ref{eq:RecursionRelation}) and the recursion based on the pole structure of the Virasoro blocks \cite{ZamolodchikovRecursion}, or the `monodromy method' \cite{Painleve6} that we reviewed and utilized in \cite{Fitzpatrick:2014vua}.  

Quantum mechanical unitarity requires that the thermal behavior of the large $c$ Virasoro blocks cannot persist at finite central charge \cite{Maldacena:2001kr}.  Recent work \cite{Maldacena:2015iua} has provided a categorization of the singularities that can occur in 4-pt correlators of two-dimensional CFTs at finite $c$ and with finite external dimensions.  The Euclidean-time periodicity of the Virasoro vacuum block of equation (\ref{eq:Vformula}) produces an infinite sequence of singularities at $z = 1 - e^{\frac{n}{T_H}}$ for integers $n$, and these should not persist at finite $c$ and $h_H$.  It would be interesting to understand how they are eliminated from the exact correlator, and how large it grows in their vicinity.

One of our motivations for this work was to search for a method of computation that might generalize to higher dimensional CFTs.  Both the recursion relation and the decomposition of the Virasoro block into CFT data are suggestive of patterns that may continue in the much-less-stringently-constrained higher dimensional CFTs.  It seems plausible that in a host of theories, the universality of CFT thermodynamics can be understood in terms of the universality of black hole geometry.  If so, this has sharp implications for higher dimensional CFT data.  This is a topic we plan to revisit in future work.

\section*{Acknowledgments}

We would like to thank Chris Brust, Cliff Cheung, Ethan Dyer, Simeon Hellerman, Shamit Kachru, Ami Katz, Daliang Li, Zuhair Khandker, Eric Perlmutter, Steve Shenker, and David Simmons-Duffin for valuable discussions.  JK and JW are supported in part by NSF grants PHY-1316665, PHY-1454083, and by a Sloan Foundation fellowship.  ALF was partially supported by ERC grant BSMOXFORD no. 228169. MTW was supported by DOE grant DE-SC0010025.  We would like to thank the Weizmann Institute and the Simons Center at Stony Brook for hospitality while this work was completed.  This work was performed in part at the Aspen Center for Physics, which is supported by National Science Foundation grant PHY-1066293.

\appendix

\section{Detailed Derivation of the Rules}
\label{sec:AppendixDetailsDerivationRules}

In this appendix we provide the detailed algebraic justification of the diagrammatic rules outlined in section \ref{sec:DiagrammaticRules}. Our derivation of these rules is inductive, demonstrating that $k$-graviton exchange automatically inherits the diagrammatic structure of the $k-1$ graviton case.

We defined a pure $k$-graviton operator with indices $a_1, \cdots, a_k$ in equation (\ref{eq:DefX}).
The Virasoro matrix elements in that definition can be evaluated to give
\be
X_{\{ a_k \}}  
= X_{\{ a_{k-1} \}} L_{a_k}  - \fr{1}{\sqrt{c}} \sum_{j=1}^{k-1} \frac{(a_j + 2 a_k)a_j(a_j^2-1)}{(a_j+a_k)((a_j+a_k)^2-1)} X_{\{ a_{k-1},\hat{a}_j, a_k+a_j \}}.
\ee
Let us then define $V_{\{ a_{k} \}}$ as the term linear in $h_L$ in the matrix element 
\begin{equation*}
\< X_{\{ a_{k} \}} \CO_L(0) \CO_L(z) \>.
\end{equation*}
Our recursive definition of the $X$ operators along with the formulas in section \ref{sec:ApproximatingV} for the action of $L_{m}$ on primary operators and on other Virasoro generators leads to the relation
\be
V_{\{ a_{k} \}}= \fr{z^{a_k}}{\sqrt{c}} \left( (s-a_k) V_{\{ a_{k-1} \}}- \sum_{j=1}^{k-1}  \frac{(a_j + 2 a_k)a_j(a_j^2-1)}{(a_j+a_k)((a_j+a_k)^2-1)}  V_{\{ a_{k-1}, \hat{a}_j, a_k+a_j \}} \right), 
\ee
where we have exclusively kept terms linear in $h_L$ and defined $s \equiv \sum_{i=1}^k a_i$.  
We will now rewrite this expression to define $V_{\{ a_k \}}$ in terms of a single $V_{\{ a_{k-1} \}}$, which already satisfies the diagrammatic rules, and then write that recursive expression itself as a sum over Virasoro diagrams. This is our inductive hypothesis.

For any single diagram $D_{\{ a_{k-1} \}}^\alpha$ contributing to $k-1$ graviton exchange (we introduce an extra label $\alpha$ to distinguish between different diagrams) we can write a corresponding formula for $D_{\{ a_{k-1}, \hat{a}_j, a_k+a_j \}}^\alpha$.  To do this, we need to replace all of the propagators running from the $j$th `leaf' to the `trunk' of the tree, as well as shift some of the overall prefactors in equation (\ref{eq:OverallFactorinRules}).  This results in the relation
\begin{equation*}
D_{\{ a_{k-1}, \hat{a}_j, a_k+a_j \}}^\alpha = D_{\{ a_{k-1} \}}^\alpha \frac{(a_j+a_k-1)(a_j+a_k)(s-a_k)(s-a_k+1)}{ a_j (a_j-1)s(s+1)}  \prod_{U_\alpha \supset j} \frac{s_{a_U} + 1}{s_{a_U} + a_k + 1}.
\end{equation*}
The product runs over all branches encountered as we follow the $j$th leaf up towards the trunk of the Virasoro diagram $D^\alpha$; we label this collection of branches with $U_\alpha$ to indicate that the associated set of propagators depends on the topology of the diagram in question.  The symbol $s_{a_U}$ denotes sums of the `momenta' $a_i$ with $i \in U_\alpha$, so the factors of $s_{a_U}+1$ correspond to propagators in the diagrammatic rules, which must be shifted to account for the replacement $a_j \ra a_k + a_j$.  We have specifically chosen the symbol `$s$' as it is suggestive of Mandelstam invariants in Feynman propagators.

Using these relations, we can write the $k$-graviton matrix element $V_{\{ a_k \}}$ as a sum over diagram topologies
\begin{equation*}
\begin{split}
&\sum_\alpha \frac{(s-a_k)(s-a_k+1)}{ s(s+1)} D_{\{ a_{k-1} \}}^\alpha \left[ \frac{s(s+1)}{s-a_k + 1}  - \sum_{j=1}^{k-1} \fr{(a_j + 2 a_k)(a_j+1)}{s_{jk}+1}
\prod_{U_\alpha \supset j} \frac{s_{a_U} + 1}{s_{a_U} + a_k + 1} \right] \\
&= \frac{(s-a_k)(s-a_k+1)}{ s(s+1)}  \sum_\alpha D_{\{ a_{k-1} \}}^\alpha \left[ \frac{s(s+1)}{s-a_k + 1}  - \sum_{j=1}^{k-1} \left( s_{jk} - \frac{a_k(a_k-1)}{s_{jk} + 1} \right) 
\prod_{U_\alpha \supset j} \frac{s_{a_U} + 1}{s_{a_U} + a_k + 1} \right],
\end{split}
\label{eq:MasterXFormula}
\end{equation*}
where we've suppressed the overall factor of $z^{a_k}/\sqrt{c}$.

Let's look at this formula for a particular diagram in the sum, i.e.\ for a particular $\alpha$.  It will have some particular tree structure, where the set of $k-1$ gravitons repeatedly bifurcates as we travel from the `trunk' to the `leaves', until we are left with single gravitons labeled with $a_1, \cdots a_{k-1}$.  We can write the contributions associated to this single diagram as
\begin{equation*}
D_{\{ a_{k-1} \}}^\alpha \left[ \frac{s(s+1)}{s-a_k + 1} - \sum_j s_{jk} \prod_{U_\alpha \supset j} \left( \frac{s_{U_\alpha} + 1}{s_{U_\alpha, k} + 1} \right)   + \sum_{j=1}^{k-1} \left( \frac{a_k(a_k-1)}{s_{jk} + 1}  \right)  \prod_{U_\alpha \supset j} \left( \frac{s_{U_\alpha} + 1}{s_{U_\alpha, k} + 1} \right) \right].
\end{equation*}
The third term is exactly what we expect for the diagrams associated to the Virasoro rules where the new $k$th graviton is attached to one of the original leaves, as pictured on the far right of figure \ref{fig:VirasoroDiagramStructures}. We can then separate it out, leaving us with
\begin{equation*}
D_{\{ a_{k-1} \}}^\alpha \left[ \frac{s(s+1)}{s-a_k + 1} - \sum_{j=1}^{k-1} s_{jk} \prod_{U_\alpha \supset j} \frac{s_{U_\alpha} + 1}{s_{U_\alpha} + a_k + 1}  \right].
\end{equation*}
Note that we can rewrite the first expression as
\be
\frac{s(s+1)}{s-a_k + 1} = \frac{a_k(a_k-1)}{s-a_k + 1} + s + a_k,
\ee
so that we obtain
\begin{equation*}
D_{\{ a_{k-1} \}}^\alpha \left[ \frac{a_k(a_k-1)}{s-a_k + 1}  +s + a_k
 - \sum_{j=1}^{k-1} s_{jk}  \prod_{U_\alpha \supset j} \frac{s_{U_\alpha} + 1}{s_{U_\alpha} + a_k + 1}    \right].
\end{equation*}
We recognize the first term as the result of attaching the $k$th graviton directly to the trunk, as pictured on the left in figure \ref{fig:VirasoroDiagramStructures}.  We now have the remaining terms
\be
R^\alpha_k = D_{\{ a_{k-1} \}}^\alpha \left[  s + a_k
 - \sum_{j=1}^{k-1} s_{jk} \left( \prod_{S_\alpha \supset j} \frac{s_{U_\alpha} + 1}{s_{U_\alpha} + a_k + 1}  \right)  \right].
 \label{eq:RemainingTerms}
\ee

We will show that these terms can be transformed into the sum of Virasoro diagrams where the $k$th graviton joins a cluster of other gravitons.  In other words, these represent diagrams where the $k$th graviton attaches neither to the `leaves' of the tree, nor directly to the `trunk' of the tree, but somewhere in the `branches'.    A generic diagram of this type is pictured in the center of figure \ref{fig:VirasoroDiagramStructures}.  

The key identity that we need for this proof is
\bq
\label{eq:KeyIdentity}
\prod_{i=1}^n \frac{s_i}{s_i + a_k} 
= 1 - \frac{a_k}{s_n+a_k} - \frac{a_k s_n}{(s_{n-1}+a_k)(s_n+a_k)}
- \cdots - \frac{a_k s_2 \cdots s_n}{(s_1 + a_k) \cdots (s_n + a_k)}.
\eq
This identity can be easily proven by induction; simply multiply by $\frac{s_{n+1}}{s_{n+1} + a_k}$ and simplify.  Notice that the terms on the right-hand side of eq.\ (\ref{eq:KeyIdentity}) are very close to what we need to obtain a $k$-graviton Virasoro diagram from $D^\alpha_{\{ a_{k-1} \}}$.  The products in the numerator eliminate a set of propagators and replace them with altered propagators, accounting for the addition of $a_k$.  The numerators in eq.\ (\ref{eq:RemainingTerms}) therefore cancel propagators of $D^\alpha_{\{ a_{k-1} \}}$ and replace them with propagators of various potential $k$-graviton diagrams.

Before describing the general case, it is useful to first see an example.  
A diagram with $k=5$ gravitons has most of the features of the general case:
\begin{equation*}
\begin{split}
\fr{R_5^\alpha}{D^\alpha_{\{a_4\}}} &= 
s+a_5 - \left( \left( s_{15} \cdot \frac{s_{12}+1}{s_{125}+1} + s_{25} \cdot \frac{s_{12}+1}{s_{125}+1}+ s_{35} \right) 
\frac{s_{123}+1}{s_{1235}+1}+ s_{45} \right)\\
&= s_{1235} - (s_{125}+a_5) \left(1 - \frac{a_5}{s_{1235}+1} - \frac{a_5 (s_{123}+1)}{(s_{1235}+1)(s_{125}+1)} \right) - s_{35} \left( 1 - \frac{a_5}{s_{1235}+1} \right) \\
&= ((s_{125} +1) + (a_5-1)) \frac{a_5 (s_{123}+1)}{(s_{1235}+1)(s_{125}+1)} + (s_{1235} + 2 a_5) \frac{a_5}{s_{1235}+1}  -2a_5  \\
&=\frac{a_5(a_5-1) (s_{123}+1)}{(s_{1235}+1)(s_{125}+1)} + a_5 \left(1 - \frac{a_5}{s_{1235}+1} \right) + (s_{1235} + 2 a_5) \frac{a_5}{s_{1235}+1}  -2a_5 \\
&= \frac{a_5(a_5-1) (s_{123}+1)}{(s_{1235}+1)(s_{125}+1)}  + ((s_{1235}+1) + (a_5-1)) \frac{a_5}{s_{1235}+1}  - a_5 \\
&=  \frac{a_5(a_5-1)}{s_{1235}+1} + \frac{a_5 (a_5-1) (s_{123}+1)}{(s_{1235}+1)(s_{125}+1)}.
\end{split}
\end{equation*}
In the second line we used our key identity (\ref{eq:KeyIdentity}), and in the third line we then arranged terms in association with each type of Virasoro diagram.  In the last few steps, we began with the term with the most propagators and rearranged it into its final form, plus a remainder that simplifies the second diagram into its final form.

The general case follows the pattern of this example.  Substituting the identity (\ref{eq:KeyIdentity}) into eq.\ (\ref{eq:RemainingTerms}), we obtain the result,
\bq
\label{eq:GeneralRemainderTerms}
\begin{split}
\fr{R^\alpha_k}{D^\alpha_{\{a_k\}}} &= s + a_k - \sum_j s_{jk} \left( \prod_{S_\alpha \supset j} \frac{s_{S_\alpha} + 1}{s_{S_\alpha} + a_k + 1}  \right) \\
&= (3-k) a_k + \sum_j s_{jk} \left( \frac{a_k}{s_{n_j}+a_k} + \cdots + \frac{a_k s_{2_j} \cdots s_{n_j}}{(s_{1_j} + a_k) \cdots (s_{n_j} + a_k)}\right).
\end{split}
\eq
We have indexed the various sums of momenta with the subscript $j$, because the product of propagators that appears depends on both the structure of the specific diagram $D^\alpha_{\{ a_{k-1} \}}$ and the choice of $j$.  Specifically, to determine which propagators appear we must start with the $j$th `leaf' of the tree $D^\alpha_{\{ a_{k-1} \}}$ and follow it up to the `trunk'.  In the second line, we have dropped the explicit factors of `$+ \, 1$', incorporating them into the $s_{i_j}$ to make the notation more compact.

Propagators near the trunk will collect together many different leaves, so they will appear as summands for many different values of $j$.  Conversely, propagators that combine only a few leaves will only appear when $j$ corresponds with one of those leaves.  The most specialized terms will tend to have the largest numbers of numerator and denominator factors, ending in some $s_{1_j}$ that collects together a group of only $2$ gravitons.  Such a term will appear as
\be
\label{eq:RecursiveTelescoping}
((s_{1_j} + a_k) + (a_k-1)) \frac{a_k s_{2_j} \cdots s_{n_j}}{(s_{1_j} + a_k) \cdots (s_{n_j} + a_k)},
\ee
where the coefficient comes from the sum of the $a_j + a_k$ factors associated with the 2 gravitons in $s_{1_j}$.  The $a_k-1$ part can be immediately identified as a Virasoro diagram, while the $(s_{1_j} + a_k)$ term cancels a propagator in the denominator.  We can then drop the completed diagram, but include the terms with the next most propagators, giving
\bq
\begin{split}
&\frac{a_k s_{2_j} \cdots s_{n_j}}{(s_{2_j} + a_k) \cdots (s_{n_j} + a_k)} + (s_{2_j} + 3 a_k-1) \frac{a_k s_{3_j} \cdots s_{n_j}}{(s_{2_j} + a_k) \cdots (s_{n_j} + a_k)} \\
&= (2s_{2_j} + 3 a_k-1) \frac{a_k s_{3_j} \cdots s_{n_j}}{(s_{2_j} + a_k) \cdots (s_{n_j} + a_k)} \\
&= (2(s_{2_j} +a_k)+ (a_k-1)) \frac{a_k s_{3_j} \cdots s_{n_j}}{(s_{2_j} + a_k) \cdots (s_{n_j} + a_k)}.
\end{split}
\eq
We have eaten up a propagator, identified a Virasoro diagram, and returned the resulting expression to a form similar to eq.\ (\ref{eq:RecursiveTelescoping}), albeit with an extra factor of 2.  This process repeats until we have proceeded all the way up the tree, identifying each term as a Virasoro diagram.  At each stage, the coefficient for $s_{i_j}+a_k$ increases by 1, until the process terminates with an overall additive factor of $(k-3)a_k$, which perfectly cancels the $(3-k)a_k$ in eq.\ (\ref{eq:GeneralRemainderTerms}). We therefore see that $R^\alpha_k$ precisely corresponds to the remaining Virasoro diagrams with the $k$th graviton attached to intermediate branches.

We have now shown that if the exchange of $k-1$ gravitons satisfies our Virasoro rules, then $k$-graviton exchange does, as well. We successfully used these rules to derive the 2-graviton contribution in eq.\ (\ref{eq:2GravitonSimplification}), which therefore means that the full semi-classical Virasoro block can be derived from these diagrammatic rules.

\bibliographystyle{utphys}
\bibliography{VirasoroRulesBib}

\end{document}